\documentclass[twocolumn]{aa}
\usepackage{graphicx}
\begin{document}

   \title{Defining and cataloging exoplanets:\\
The exoplanet.eu database}

   \subtitle{}

   \author{
          J. Schneider\inst{1}%\fnmsep\thanks{Just to show the usage
          %of the elements in the author field} 
         \and C. Dedieu\inst{2}   \and P. Le Sidaner\inst{2}  \and R. Savalle\inst{2} \and I. Zolotukhin
         \inst{2} \inst{3} \inst{4} }

   \offprints{J. Schneider}

   \institute{Observatoire de Paris, LUTh-CNRS, UMR 8102, 92190, Meudon, France\\
             \email{Jean.Schneider@obspm.fr}
             %\thanks{The university of heaven temporarily does not accept e-mails} \and \institute{Cornell}
             \and Observatoire de Paris, Division Informatique de l'Observatoire, VO-Paris Data Centre, UMS2201 CNRS/INSU,  France
             \and Observatoire de Paris, LERMA, UMR~8112, 61 Av. de l'Observatoire,
             75014 Paris, France
             \and Sternberg Astronomical Institute, Moscow State University, 13
             Universitetsky prospect, Moscow, 119992, Russia}

   \date{Received ; accepted }

   \abstract{We describe an online database for extra-solar planetary-mass
candidates,
updated regularly as new data are available. We first discuss 
criteria for the inclusion of objects in the catalog: "definition" of a planet and several
aspects of the confidence level of planet candidates. {\bf We are led to point out the conflict between 
sharpness of  belonging or not to a catalogue and fuzziness of the confidence level.}
We then describe the different tables  of extra-solar planetary systems, including unconfirmed candidates
(which will ultimately be confirmed, or not, by direct imaging). It also provides
online tools: histogrammes of planet  and host star  data, cross-correlations 
between these parameters and some VO services. Future evolutions of the database are presented.
   }
\authorrunning{Schneider et al.}
 \titlerunning{The exoplanet.eu database}

   \maketitle
%
%___________________________________________________________________

\section{Introduction}

The study of  extrasolar planetary systems has become 
a very active field which will grow continuously in the coming years and
decades. This new field of astronomy leads to two types of activities:
the detection of new planets and 
new observations of known planets and on the other hand
the understanding of  physical and dynamical
processes of  individual planets,  planetary systems 
and  interactions of planets with their host stars. 

These activities  require a precise knowledge of  the characteristics
of planets and of their parent stars, i.e. a well documented catalogue. Exoplanetology is 
developing so rapidly (and this evolution will even accelerate in the coming
years) that any static catalogue\footnote{Like the table of stellar and
planet parameters by Fischer \& Valenti (2005).} is obsolete
on time scales of a few  months. 
An evolutionary online catalogue through the Internet
is better adapted to that situation.
It has the advantage of updating permanently
the data and of making possible online tools for their pre-treatments. 
Here we describe a freely accessible database consisting in a catalogue and associated online services. 
In section 2 we describe the purpose and philosophy of the database. In section 3 we discuss 
criteria for the inclusion of objects in the catalogue. In section 4 we give the detailed
content of the catalogue. In section 5 we describe the associated online services.
The database is part of a wider portal, the Extrasolar Planets Encyclopaedia, offering other services which we describe shortly in section 6. We end by
sketching  future developments in section 7.
\section{Purpose of the catalogue}
A first online catalogue, the Extrasolar Planets Catalogue, was established,
at the dawn of the Web, in 
February 1995 after the suspicion of the discovery of $\gamma$ Cep b 
(Campbell et al. 1988) and the
confirmation of the first pulsar planets 
(Wolszczan 1994)\footnote{Some ancient pages, back to 1999,  are available at \\
{\rm http://web.archive.org/web/*/http://www.obspm.fr/planets}
{\bf AND AT {\rm http://WEB.ARCHIVE.ORG/WEB/*/HTTP://EXOPLANET.EU}}.}.
It was followed some years 
later by the California and Carnegie 
Planet Search table at {\rm http://exoplanets.org/planet\_table.shtml} (Butler et al., 2006) and
by the Geneva Extrasolar Planet Search Programmes  Table at 
{\rm http://obswww.unige.ch/}$\sim${\rm naef/planet/geneva\_planets.html} (Mayor, 
Queloz, Udry and 
Naef). The Extrasolar Planets Catalogue,  available since then at
{\rm http://exoplanet.eu/catalog.php} (Martinache \& Schneider 2004), has been upgraded in 2005 by
the addition of several graphical and statistical online services (Le Sidaner et al. 2007). 
This new version has been followed by two
other professional online databases:   the NStED Database at 
{\rm http://nsted.ipac.caltech.edu}  and the Exoplanet Data Explorer at
{\rm http://exoplanets.org} (Wright et al 2010), the later providing some online tools.
 The Exoplanet Data Explorer and the Geneva catalogue have the advantage of providing,
for some planets, first hand data
by the observers (and often discoverers) themselves. But as of February 2011
these catalogues list only planets discovered by radial velocity and by transits
whereas our catalogue also lists planets discovered by astrometry\footnote{Although 
there is, as of February 2011, no planet discovered by astrometry; indeed, the planet VB 10 b 
claimed to be discovered by astrometry (Pravdo \& Shakland 2009) 
has not been confirmed by radial velocity measurements (Bean et al. 2010).},
direct imaging, microlensing and timing. It also gives a table of unconfirmed 
or problematic planets (see sections 3 and
4 for criteria of "confirmation").
For completeness, there was also a fifth list, provided by the
IAU Working Group on Extrasolar Planets at 
{\rm http://www.dtm.ciw.edu/boss/planets.html}, but it only listed 
planet names and, as of February 2011, it is no more maintained.

 As a final introductory remark, we point out that our catalogue
includes all planets which we estimate to be ``reasonable'' candidates,
including a separate table of unconfirmed planets. 
It is aimed to be a working tool permanently in progress.
This
choice is made to provide to the community of researchers all
the available information at any time, allowing them  to make their
own judgement and to use the data for new observational or theoretical work
and to confirm, or not, problematic candidates by complementary observations.
It is also designed to help high-level amateurs (e.g. for transiting planets)
and, being updated daily, to give the latest news for  correct information 
for outreach activities.
For that latter purpose it is multi-lingual (English, Farsi (Persian), French, German, 
Italian, Polish, Portuguese and Spanish). For comparison, the Exoplanet 
Data Explorer gives only "secure" planets in the sense that they are all
published as such in refereed journals\footnote{It also provides more 
detailed planetary  and stellar data
(e.g. planet density, $V.\sin i$ for the star).}. But as a counterpart it contains less candidates. 
We have chosen to provide a larger sample of candidates since each user can make her/his own mind
on their validity.
\section{Criteria for the inclusion of objects in the catalogue}
The first task is to choose which objects to include in the catalogue.
The question is very simple, but the answer is delicate.
It faces several problems for which one has to 
make choices. It therefore deserves a  discussion to clarify all problematic aspects.
The objective of making a catalog of exoplanets rests on the implicit prejudice
that there is a well definable category of objects sharing some common nature
with Solar System planets. {\bf This purpose is necessary if the catalog
is to be used to draw statistical features of planet characteristics.}
It is thus
essential to establish criteria in order to
decide which objects deserve the name "planet". 
As we will see, the ideal situation of criteria ruling all configurations without
ambiguity cannot be realized.
\subsection{"Definition" of a planet}
The word ``definition'' refers to two different
situations. First, it means an arbitrary convention, like for instance  the neologism "pulsar".
But   the word "definition" also often designates
an attempt to clarify {\bf the content of a pre-existing word} for which we have some spontaneous 
preconceptions, whatever their grounds, and  to catch an (illusory) "essence"
of what is defined. It is then made use of pre-existing plain language words
which carry an {\it a priori} pre-scientific content likely to introduce some confusion in the reader's mind.
In the clarification of pre-scientific conception versus scientific convention, some 
arbitrariness is unavoidable.

Here we do not try to catch an essence of planets but to find pragmatic
criteria for the inclusion of objects in the catalogue based on physical
properties, if posssible, and on observable appearance.
The complexity of the problem arises when one seeks to correlate 
these two approaches.
\subsubsection{Physical nature of planets}
Until 2001, a planet was defined as not having central deuterium burning.
According to sub-stellar interior models, this criterion enables to define
a planet as having a maximum 
mass  around 13 Jupiter mass (Burrows et al. 2001).  But the
discovery {\bf of the companion HD 168443 c with $M.\sin i = 18.1  M_{Jup}$ }
to the  star HD 168443 having already {\bf a 
companion with $M.\sin i =  8.2 M_{Jup}$} (Marcy et al. 2001) introduced a complication (see section 4 for the labelling of planets). The idea emerged that
a substellar companion with a mass larger than the 13 $M_{Jup}$ limit could 
share with less than  13 $M_{Jup}$ objects a common "nature", whatever it is. 
The fact that there is no special feature around 13 $M_{Jup}$ in the observed mass spectrum 
(Fig 1 Udry 2010) reinforces the choice to forget this mass limit.
But then an embarrassing problem  arises: where to set this limit (if a limit makes sense)?

\begin{center}
\includegraphics[width=7cm]{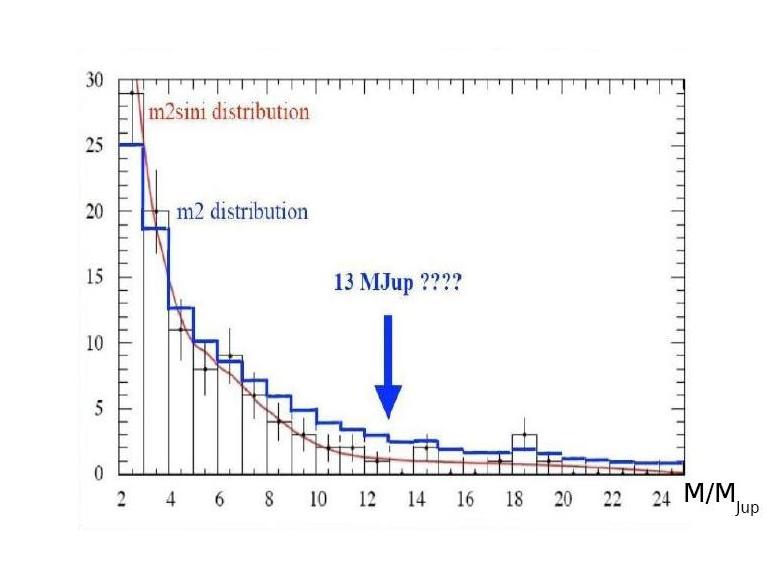}

{\bf Fig. 1.} Mass distribution of companions below 25 $M_{Jup}$ (Udry 2010)
\end{center}

Another approach is based on the formation scenario. The convention is then to
call ''planets'' objects formed by accretion of planetesimals in a circumstellar dust disc,
just by analogy with Solar System planets, and to call "brown dwarfs" objects formed 
by collapse in a gas cloud or circumstellar disc, by analogy with stars.
We have chosen to follow as much as possible this approach.
But it faces a major difficulty: for a given substellar companion, how to know
if it is formed by core accretion or by gravitational collapse since we cannot 
catch the formation process "in vivo" and it is not clear how to infer it from current observables.
 Unfortunately there is an overlap in the guessed mass
distribution of planets and brown dwarfs (Baraffe et al. 2010, Spiegel et al. 2011). 
Baraffe et al. (2010) suggest that the brown dwarf mass function can go down to 6 $M_{Jup}$.
Therefore the companion mass can {\it a priori} not serve as a 
reliable criterion for deciding if it should be named a ''planet'' 
or a ''brown dwarf''. It is likely that these two categories have
different statistical distributions for instance
in the (mass, semi-major axis) plane, but this does not
help for individual objects. 
One could make use of the bulk density of objects:
those formed by core accretion may have more heavy elements (Baraffe et al. 2010).
But then one has to know their radius.
As long as ultra-high angular resolution imaging cannot measure directly
this radius, the latter is known only for transiting planets.
In addition it is scrambled
by the observed "radius anomaly", i.e. the abnormally large observed radius
compared to models (at least for hot Jupiters, Baraffe et al. 2010).
The companion mass value is finally the only 
simple pragmatic present criterion for the designation "planet" and we chose to rely on it.
But we still have  to choose a mass limit.

There is no theoretical prediction
for the mass spectrum of brown dwarfs, but  there is 
a dip around 25-30 $M_{Jup}$ in the observed  distribution {\bf in $M.\sin i$} of substellar objects
(Udry et al. 2010, Fig. 2a). A closer look reveals a flat quasi-void between $\sim$ 25 - 45 
$M_{Jup}$ (Sahlmann et al. 2011 - Fig. 2b).  {\bf Since according to Baraffe et al. (2010) the likelihood of
an object to be a brown dwarf increases with mass, and, since the observed  mass spectrum 
       decreases from a few to 20 Jupiter mass, we attribute this decrease 
       to the mass spectrum of planets, and finally since a threshold 
       has to be chosen, we {\it arbitrarily} choose (and perhaps provisionnally 
       if new insights on the mass distribution emerge in the future)
         to priviledge at this stage a maximum mass of 25 $M_{Jup}$.}

\begin{center}

\includegraphics[width=7cm]{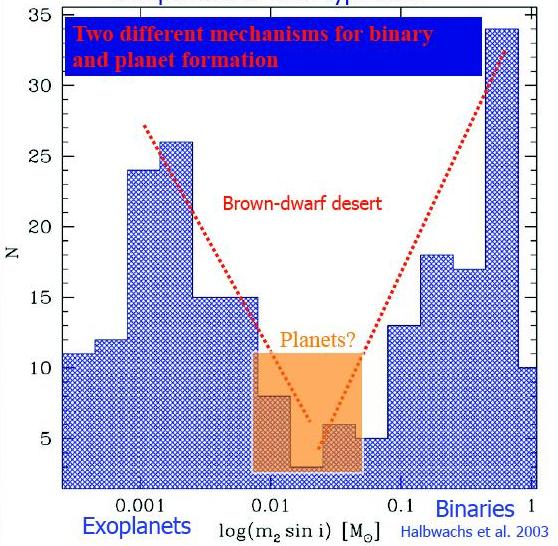}

{\bf Fig. 2a.}  Mass distribution of substellar objects (Udry 2010)

\includegraphics[width=7cm]{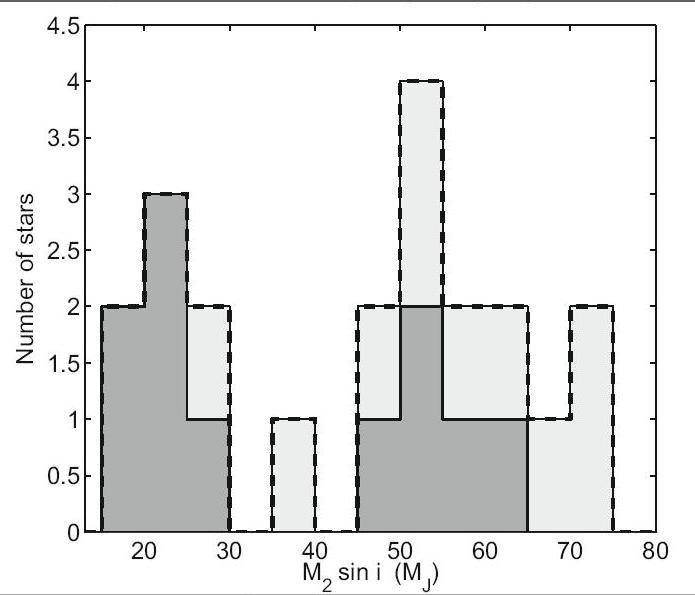}

{\bf Fig. 2b.}
Zoom of the mass distribution around 25 - 40 $M_{Jup}$ (Sahlmann et al. 2011)
\end{center}

There is also a break in the radius distribution as a function of substellar masses 
around 25 $M_{Jup}$ (Fig 3a Pont et al. 2005, 3b {\bf Anderson et al.} 2011),
suggesting a difference in physical nature below and above this mass.

\begin{center}
\includegraphics[width=7cm]{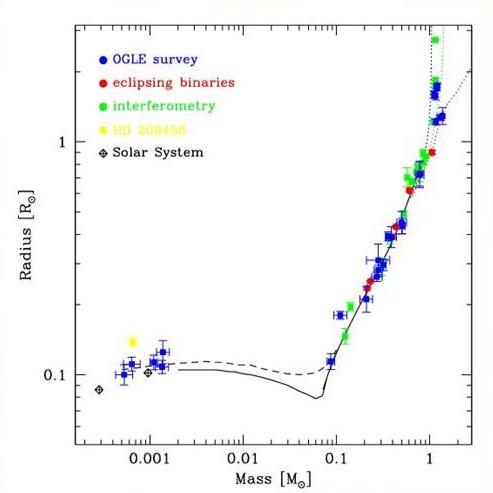}

{Fig. 3a.} Radius distribution as a function of mass for substellar objects (Pont et al. 2005)

\includegraphics[width=7cm]{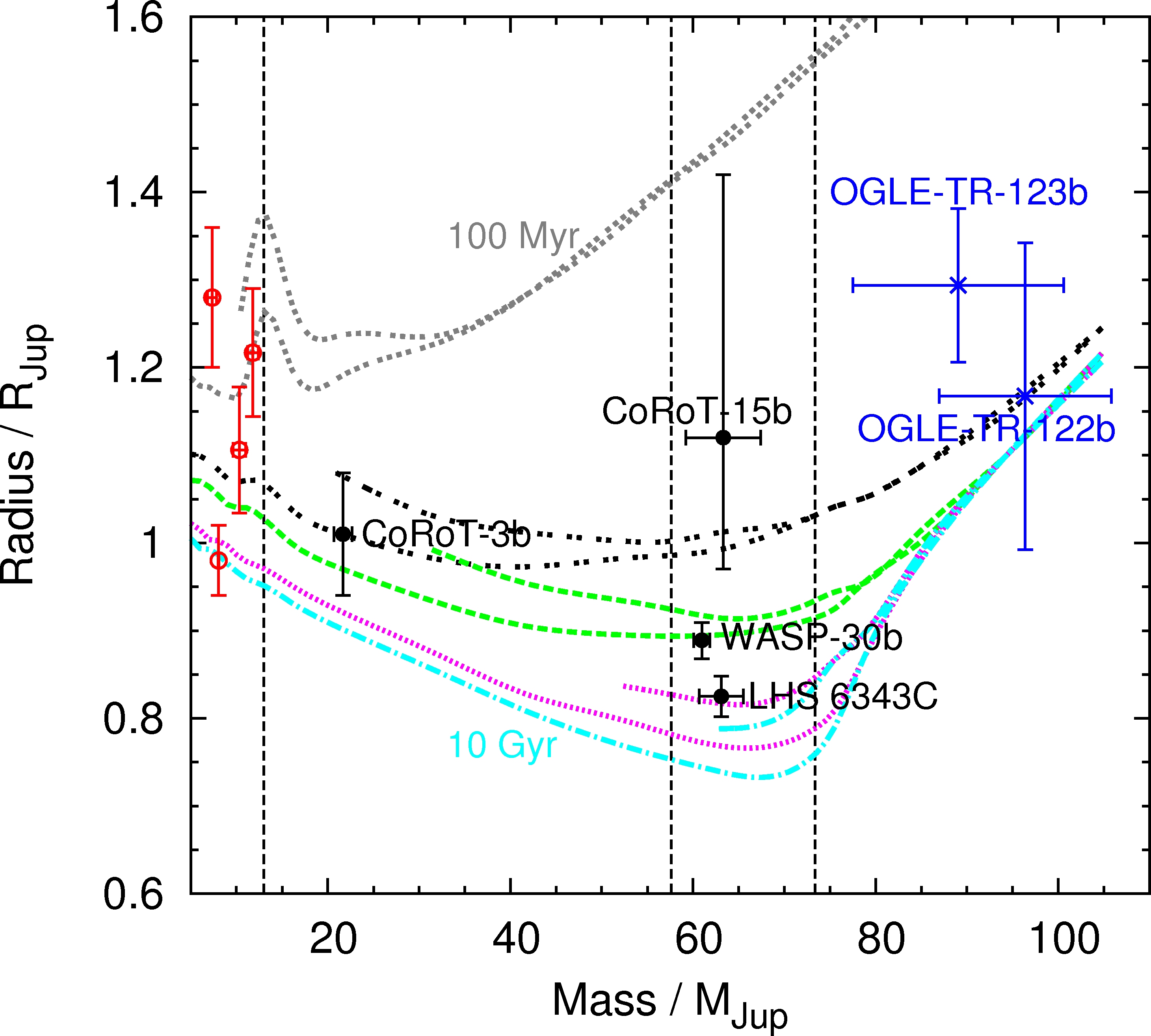}

{Fig. 3b.}  Zoom of the mass-radius distribution around 25 - 40  $M_{Jup}$ (Anderson et al. 2011).
\end{center}

These facts make plausible that the population below around 25 $M_{Jup}$
is essentially made of planets (in the above sense).
Since in the 13 - 25 $M_{Jup}$ region there are only about 5\% of all objects with a 
mass below 25 $M_{Jup}$, a possible confusion
between planets and brown dwarfs  is statistically not significant.

The word "superplanet" has sometimes been proposed for massive planets (e.g. Udry et al. 2002), 
but no clear definition was given in the literature. One could
use  it to designate objects between 13 and 25 $M_{Jup}$, but since
there is no special feature around 13 $M_{Jup}$ in the observed mass spectrum
of substellar objects (Fig. 1), it 
introduces an unnecessary complication and we  discard it.

Another property  of planets in the heliocentric view is that they are 
low mass companions of a parent
star. But a few isolated objects with a mass probably below  $\sim$ 10
Jupiter mass  have  been discovered (e.g. S Ori 70 (Zapatero-Osorio 
et al. 2002)). We therefore include an additional table of ''free-floatting'' planet candidates
(called "rogue" planets by Abbot \& Switzer (2011)).
 Since some of these objects  may be "true" planets ejected from a circumstellar
orbit, we avoid to designate them by a new word, "sub-brown-dwarfs", evoked by Spiegel \& Burrows (2011).  

The International Astronomical Union is attempting to
establish an "official" definition of exoplanets (Boss et al. 2010),
but we feel that it is  premature since the situation is still
evolving and that any definitive definition is likely to be too rigid
to adapt to new discoveries. We note that, with  the exception of Solar System bodies,
there is  no other "official" definition of categories of objects 
in astronomy (e.g. of quasars) and it is better like this.
\subsubsection{Observable parameters}
A growing activity is and will be the spectro-polarimetric characterization of substellar companion atmospheres.
The physics of atmospheres of gas giant planets and of brown dwarfs is essentially the same. 
Some users of the database are more interested by this aspect than by the internal 
structure resulting from a formation scenario. It is therefore 
useful for them to have a database including some brown dwarfs. It is  
another reason to have a ''generous'' upper mass limit of 25 $M_{Jup}$.

{\bf Another concern is a lower mass limit for the planet: it is not so academic
since objects transiting a white dwarf with  a size  
$0.1 R_{earth} \approx 0.25 R_{pluto}$ 
or companions to planets detected by transit time variations
with a mass smaller than the moon (Sartoretti \& Schneider 1999)
may be detected any time in the nearby future. It is not sure that
the IAU resolution B5 in 2006 defining a planet as "a nearly round
object which has cleared the neighbourhood around its orbit"
will be pertinent for exoplanets. For the present catalogue we will not
use any lower mass or size limit (and therefore include future "exo-moons"). 
We incidentally note that the planetary nature of PSR 1257+12b ($M = 1.8 M_{Moon}$)
has never been contested in the litterature.}

As a provisional conclusion  we therefore chose at this point to include all objects 
with a mass below 25 $M_{Jup}$.  One cannot exclude that this limit will change in the future.
Note that a sharp limit is somewhat absurd, since in the present case it
excludes 25.1 $M_{Jup}$ companions, but at the same time we see no pragmatic alternative to sharpness.
\subsection{Confidence level of a planet}
Once we have chosen a methodology to decide which objects are called planets, 
we have to decide whether a candidate fulfills these criteria or not.

There are two aspects:

- is the suspected companion real or an artefact (stellar activity etc)?

- is the companion, if confirmed, really a planet? 

{\bf The ideal situation would be that authors give a clear confidence level for each 
candidate and the task of a compiler would be to decide (arbitrarily) 
 a threshold for a confidence level. But in most paper there is no clear confidence
level, making
the decision of confirmed/unconfirmed  arbitray and subjective. }
\subsubsection{Companion or artefact?}
There is first a well known problem of confidence level of the interpretation  of observations,
specific to each  detection method. After the analysis of artefacts, specific to each method,
the planet candidate has a False Alarm Probability (FAP). The California Planet Survey team 
requires systematically a FAP below 5\% to announce a planet discovery (Howard et al. 2010). 
But all authors do not give a clear estimate of their FAP 
(sometimes difficult to evaluate quantitatively) and in this respect the 
confidence level of planets is heterogeneous in the catalogue. 
We therefore  adopt, 
with only one exception (Gl 581 g - see special cases below), the criterion of being
published in a refereed paper or presented in a professional conference or
website as
a "secure" companion.
Without entering a detailed discussion of all artefacts,
let us remind some of their  features:

-  {\it  Radial velocity (RV)}\\
An apparent RV variation can be due to various phenomena in the stellar atmosphere.
They are generally eliminated by the "bisector variation" test {\bf (Mandushev et al. 2005)}.

Once the stellar wobble has been confirmed, it could still be due to another mechanism
than the pull by a companion.
As shown by Schneider (1999) and confirmed by Morais \& Correia (2011), 
it can be due to a distant binary star with the same orbital period
as for the planet (false) candidate. But such events can mimic
only  low mass planets on wide orbits, beyond the present detection capability. 
It does not apply to objects currently present in the catalogue.
It nevertheless represents  a danger for future discoveries with the
Gaia astrometric space mission (Schneider \& Cabrera 2006).

The final discrimination between this artefact and a real companion
will be provided by the detection of the candidate by direct spectro-imaging.
The same discussion holds for planets discovered by astrometry and by timing.

- {\it Transits}

The main possible artefact is the presence of a background eclipsing binary (BEB) in the target 
Point Spread Function. A first test is then to check, if possible, that the transit is 
{\bf at first sight} achromatic
(we leave aside the discussion of self-luminous transiting planets)
and that a BEB is not seen in high angular resolution images.
A second step is then to measure the radial velocity variations
due to the suspected transiting object {\bf (after elimination of the bisector variation 
(Mandushev et al. 2005))}. 

{\bf We note that new approaches for the consolidation of the consolidation of the planetary nature 
of the transiting have recently emerged: indirect measurement of the object mass by
Transit Time Variation as for Kepler-11b,...,g (Lissauer et al. 2011), 
or the highly accurate achromaticity of the transit as for Kepler-10c (Fressin et al. 2001).}

Once the transit has been secured, it can still mimic a Jupiter-sized planet if it is 
due to a brown dwarf,
or a super-Earth if it is due to a white dwarf. {\bf This case} is removed by RV measurements
which give the companion mass.

- {\it Direct imaging}

The main possible artefact is the confusion with a background star. It can easily be removed
by verifying that the star and the companion have a common proper motion. 
Some authors nevertheless publish
a planetary candidate "to be confirmed", prior to the check of common proper motion,
 when the probability of presence of a background star
in the target vicinity is "sufficiently" low (e.g. Lagrange et al. (2008) for $\beta$ Pic b). 
There is no commonly accepted value of the FAP  for these cases
and the decision between "confirmed" and "unconfirmed" planet is arbitrary.
\subsubsection{Planetary nature of the companion}
Here again, there are two aspects: 

- {\it Is the companion mass below the chosen 25 $M_{Jup}$ limit?}\\
Once the mass limit has been decided,
an uncertainty remains since the measured mass suffers from 
various inaccuracies. It introduces a fuzziness on what objects to include in the catalogue.
For all type of methods there is an instrumental uncertainty. 

But in case of planets 
detected by RV or by timing  there is an additional "$\sin i$ uncertainty", meaning that the only observable
is then $M_{pl}.\sin i$ instead of $M_{pl}$ itself, due to the unknown orbital inclination $i$.
The database lists all objects with $M.\sin i$ less than 25 $M_{Jup}$.
As it is well know, this uncertainty does not hold for transiting planets since 
$i$ is then derived from the shape of the transit light curve.
For planets detected by imaging, constraints on their mass can be put thanks
to multicolor photometry or spectroscopy and to atmosphere modelling.
In that case the uncertainty is rather large (see for instance Neuhaeuser et al 2005).
The mass estimate then rests most often on the Baraffe et al. model (2010) 
correlating the mass to the spectrum and age.
It is important to note that
this model has not been
tested yet for planets with a mass known from radial velocity measurements.

- {\it Is the  less than 25 $M_{Jup}$ companion a planet?}\\
For rigorous completeness one has to consider the possibility that less than 25 $M_{Jup}$
compact objects are not planets. Presently there is no known category of such objects. But,
at least in the early era of the first  planet candidates when the 
abundance of exoplanets was completely unknown, one could wonder if a new class 
of objects was not discovered instead. For instance one could in principle
invoke black holes or "X-stars", analogous to neutron stars, where X is some new type of particle.
Planetary mass black holes have an evaporation time $\tau _{BH} = \tau_{Planck}(M_{BH}/M_{Planck})^3 \approx 10^{63}(M_{BH}/M_{Jup})^3$s (Yang \& Chang 2009) and therefore survive evaporation over stellar lifetimes.
But,
having a (Scharzshild) radius of the the order of meters, they are made implausible 
by the observed radius derived from transit events. 
X-stars would have a mass $M_{X*} = M_{Planck}^3/M_X^2$, requiring a 30 neutron mass X particle
for a $1 M_{Jup}$ object. It is not excluded by the zoo of "beyond the standard model" theories,
but it would imply a discrete mass distribution with only a very few mass values (one per X-species).
These perspectives are (or were) an important issue given the philosophical significance of exoplanets
and that "extraordinary claims require extraordinary proofs". But of course today the case is closed.

In conclusion, we have arbitrarily chosen a 25 $M_{Jup}$ + $\Delta M$mass limit with a 1 sigma
margin for $\Delta M$. By having a "generous" mass limit, we allow the user to compare easily planets and brown dwarfs.
Here again a sharp mass limit, even with a 1 sigma margin, is absurd since it excludes companions
with  $M$ - $\Delta M$ = 25.1 $M_{Jup}$.
\subsection{Ambiguities and uncorrect characteristics attribution}
Some ambiguities are present in the interpretation of stellar wobbles.
Anglada-Escude et al. (2010) have warned that eccentric orbital solutions can hide
two planets in a 2:1 resonance (and vice versa?).
Other degeneracies are present for two planet systems: 
 exchange orbits  (change in semi-major axis, 
Funk et al. 2011), eccentric resonances (change in eccentricity, Laughlin \& Chambers 2002, Nauenberg 2002),
 Trojan planets (Dvorak et al. 2004), 
large moons or binary planets (Cabrera \& Schneider 2007).
These ambiguities
will finally be resolved by the detection of the candidate by direct spectro-imaging.

For direct imaging,  Kalas (2008) and  Kennedy \& Wyatt (2011) have pointed out that
the planet can be surrounded by a large dust cloud leading to a significant overestimate 
of its radius and albedo.
\subsection{Special cases}
Some special cases deserve a few comments:

 - {\it Planets designated as "brown dwarfs"}\\
Some authors designate their discovered substellar companions as "brown dwarfs" whereas
they have a mass below the 25 $M_{Jup}$ limit. It is for instance the case of HIP 78530 b 
($M$ = 23.04 $\pm$ 4 $M_{Jup}$ (Lafreni\`ere et al. 2011). We have included these allegated brown 
dwarfs in the main planet table.

- {\it Gliese 581 g}\\
The individual case of Gliese 581 g has deserved much attention because, as a one of the first
potentially
habitable planets, it is emblematic. It has been published in a refereed paper (Vogt et al. 2010)
and as such 
should normally be in the main table. It has been challenged by Pepe et al. (2010) and
by Gregory (2011), but as of February 2011 with no published additional RV data. We have chosen
to transfer it (provisionnally?) in the table of unconfirmed planets.

- {\it Objects with very high mass uncertainty}\\
 Some objects  have a published mass well beyond the 25 $M_{Jup}$ mass limit, but with
a very large mass uncertainty $\Delta M$ so that the value for $M-\Delta M$
is below the 25 $M_{Jup}$ limit. It is for instance the case of HD 190228 b 
for which Reffert \& Quirrenbach (2011) give a mass range 5.93 - 147.2 $M_{Jup}$ at the 3 $\sigma$ level
by using Hipparcos astrometric data.
We have  transfered them into the table of unconfirmed planets. The situation will 
be clarified around 2015 with the results of the ESA Gaia astrometric mission.

{\bf  Some objects have completely unknown mass. We then use another criterion,
the  size; for some objects, like for instance SDSS J083008+4828  ($R = 0.61 R_{Jup}$
Tsuji et al. 2011), a radius based on the infrared flux has been determined. We include as unconfirmed
planets all such objects with an upper radius
provisionally set to $1.2 R_{Jup}$.}

- {\it Planets declared unconfirmed by the discoverers}\\
They are naturally in the table "unconfirmed". 
{\bf Here again some borderline cases are inevitable. It is for instance the case, as of June 2011,
for Lupus-TR-3b (see the corresponding page "Notes for Lupus-TR-3b".}

- {\it Suspected planets with no clear parameters}\\
It is the case for candidates suspected from a linear trend in RV monitoring, 
or planets suspected to sculpt a debris disc or planets suspected because they pollute a stellar spectrum.
For the two first cases, the cadidate will ultimately be confirmed, or not, by direct imging.
They are in the "unconfirmed" table.
\section{Description of the catalogue}
We describe the catalogue as it is in February 2011. It may evolve
continuously. It is, provisionally,  organized in 8 tables, according to their discovery methods.
We distinguish
"detection" from "discovery": e.g. some planets are discovered by RV and detected
by transit afterwards. In the coming years, planets discovered by RV will be detected by direct imaging
and vice-versa; therefore this categorization is likely to change. The 8 tables are:\\
1/ All "confirmed" planets\\
2/ Planets discovered by RV and/or astrometry (note that as of February 2011 no confirmed
planet has been discovered by astrometry, although a few of them have been observed 
 in an astrometric monitoring after their discovery by RV).\\
3/ A sub-table of the previous collects planets discovered first by transit and confirmed later by RV,
and planets discovered first by RV with transits discovered afterwards.\\
4/ Planets discovered by microlensing\\
5/ Planets discovered by direct imaging\\
6/ Planets discovered by timing (pulsar planets, timing of eclipses 
of eclipsing binaries (or planetary transits) or timing of stellar oscillations).\\
7/ Unconfirmed or retracted planets\\
8/ "Free floatting" planets.

For each planet the tables list the planet and parent star characteristics.
In addition there is an individual page for each planet and for each planetary system.
It is accessible by clicking on the planet name in the table (Fig 4.).

\begin{center}
\includegraphics[width=7cm]{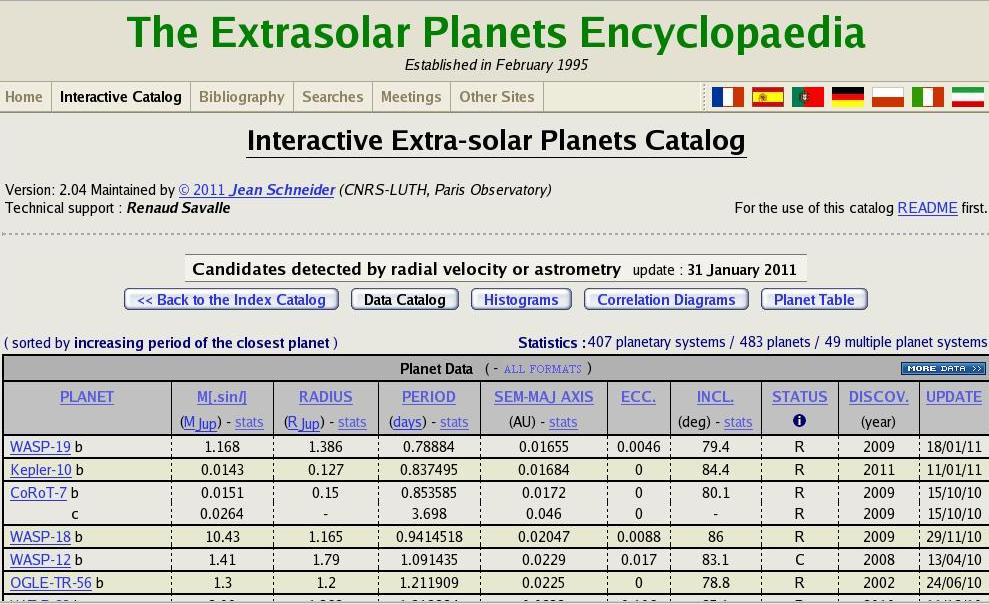}

{\bf Fig. 4.} Excerpt from the table of candidates discovered or detected by radial velocity or astrometry
\end{center}

The planet characteristics in the tables are given (when available) by Table 1.

\begin{table*}
\begin{center}
{\bf Table 1}  Planet parameters in planet tables\\

\begin{tabular}{|l|l|l|}\hline
{\bf Entity}& {\bf Designation} & {\bf Unit} \\\hline
Name &PLANET & \\
Mass & M[.sin {\it i}] & Jupiter and Earth mass\\
 &{\it (see comment below}&{\it (see comment below)}\\
Radius & RADIUS& Jupiter and Earth radius \\
     & &                {\it (see comment below)}\\
Orbital period &PERIOD&  days  and years\\
  & &                              {\it (see comment below)}\\
Semi-major axis & SEM-MAJ. AXIS & AU\\
Eccentricity & ECC. & \\
Inclination & INCL. & degrees\\
Publication status & STATUS & R, S, C, W \\
  & &                              {\it (see comment below)}\\
Year of discovery & DISCOV. & year\\
Date of data update. & UPDATE &  dd/mm/yy \\\hline
\end{tabular}
\end{center}
\end{table*}

The parent star characteristics given by individual tables 
are shown in Table 2 and Fig. 5 

\begin{center}
\includegraphics[width=7cm]{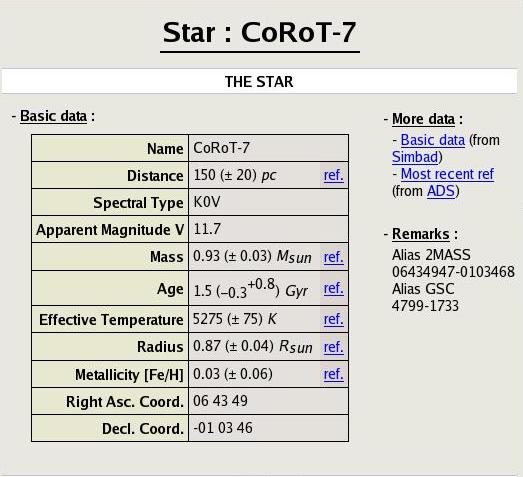}

{\bf Fig. 5.} Parent parameters given by individual tables (example of CoRoT-7)
\end{center}

\begin{center}
{\bf Table 2}  Parent star parameters\\

\end{center}

\begin{center}
\begin{tabular}{|l|l|l|}\hline
{\bf Entity}& {\bf Designation} & {\bf Unit} \\\hline
Distance & DIST. & pc \\
Spectral type & &\\
Apparent  & MAG. V & \\
magnitudes  & MAG. I & \\
V, I, H, J, K & MAG. H & \\
          & MAG. J & \\
          & MAG. K & \\
Mass &MASS & M$_{\odot}$\\
Radius & RADIUS & R$_{\odot}$\\
Metallicity& [Fe/H] & \\
Right asc. & ALPHA & hh mm ss\\
Declination & DELTA & dd mm ss\\\hline
\end{tabular}

\end{center}

\bigskip

The individual pages  for planets and planetary systems "Notes for planet xx"
contain additional details:\\
- they give the  quantities listed in Table 1 plus data listed in Table 3 
  (when available) and their errors.\\
- molecules detected in the planet atmosphere\\
- specific remarks\\
- bibliography relevant for the planet (sorted by author or publication date)\\
- links to professional websites associated with the planetary system. Data may be continuously refined on observers web pages: therefore we give a link to these
pages\\
- link to Simbad and ADS pages of the corresponding parent star.

The Fig 6 gives as an example the individual page for HD 189733 b.

\begin{center}
\includegraphics[width=7cm]{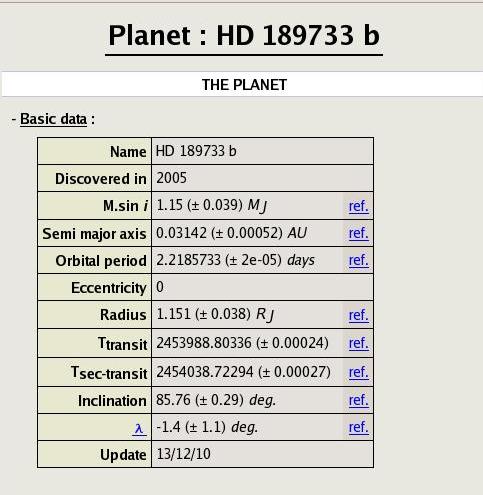}

{\bf Fig. 6.} Example of planet parameters given by individual pages for
HD 189733b
\end{center}

\begin{table*}
\begin{center}
{\bf Table 3} Additional star and planet parameters in individual pages\\
\begin{tabular}{|l|l|l|l|}\hline
{\bf Entity}& {\bf Designation} &   {\bf Unit} & {\bf Comment} \\\hline
Temperature of the star& Effective Temperature &K &\\
Longitude of periastron & $\omega$ & deg. & For eccentric orbits \\
Longitude of ascending node & $\Omega$ & deg & \\
Orbit "misalignement" & $\lambda$  & deg & For transiting planets\\
   &  &  & {\it (see  section 4.1)}\\
Epoch of transit & $T_{transit}$ & JD & For transiting planets \\
Epoch of secondary transit & $T_{sec-transit}$ & JD & For transiting planets \\
Epoch of passage at periastron & $T_{peri}$ & JD & For eccentric orbits\\
Epoch of max. star velocity & $T_{Max RV}$ & JD & For circular  orbits\\\hline
\end{tabular}
\end{center}
\end{table*}
\bigskip

For multiplanet systems there is a synthetic page with all planets in the system and their characteristics
(click on the star name in individual pages).

Links to individual planet pages of the Exoplanet Orbit Database at exoplanets.org 
are under development.
\subsection{Comments}
- {\it Source of data:}\\
Data are the latest known. They are updated daily. As of February 2011 they are taken from\\
   - Latest published papers or preprints, or conference proceedings.\\
   - First-hand updated data on professional websites. They presently are: Anglo-Australian Planet Search   
     at {\rm http://www.phys.unsw.edu.au}$\sim${\rm cgt/planet/AAPS}\_{\rm Home.html}, California  Planet Survey at 
     {\rm http://exoplanets.org/}, Geneva Extrasolar Planet Search Programmes
      at {\rm http://obswww.unige.ch/exoplanets}, 
     Kepler candidates at {\rm http://archive.stsci.edu/kepler/planet}\_{\rm candidates.html},
     SuperWASP at {\rm http://www.superwasp.org},
     and University of Texas - Dept. of Astronomy at {\rm http://www.as.utexas.edu}.

- {\it References for data:}\\
In each individual pages there is a flag "ref" for each data. By clicking on that flag
the user is directly connected  to the reference paper giving the corresponding value.
These references are updated as they appear in the literature.

- {\it Mass $M.[\sin {\it i}]$:}\\
For planets detected by radial velocity and timing, only the product $M.sin {\it i}$, where {\it i}
is the orbit inclination, is  known
in general. For transiting planets, {\it i} and therefore $M$
is known from the fitting of the transit lightcurve. 
For planets detected by astrometry {\it i} is directly infered from the parent star orbit. 
For planets detected by radial velocity in multiplanet systems, it can sometimes be infered from
the dynamical analysis  of the planet-planet
interaction (deviation from purely keplerian orbits - e.g. Correia et al. (2010) for the GJ 876 system) and in a few years it will be infered from direct imaging of some planets. {\bf The bracket in $[.\sin i]$ then means
that it has to be ignored when the inclination is known and the value for
$M.[\sin {\it i}]$ is the true mass value $M$.}

- {\it Semi-major axis $a$: }\\
When the semi-major axis $a$ is not given in a detection paper, it is
  derived from the published orbital period and from the mass of the parent star through the Kepler law
$P=2\pi \sqrt{a^3/GM_{\star}}$.

- {\it Parent star coordinates and magnitudes:} \\
If not given in a detection paper, they are  taken from SIMBAD.

- {\it Orbit "misalignement" $\lambda$:}\\
The precise definition of $\lambda$ is: the angle between the sky projections of the perpendicular to the planet orbit and of the star rotation axis (if the planet orbit is in the star equatorial plane, $\lambda$ = 0). 
It is infered from
the measurement of the Rossiter-McLaughlin effect. Some authors give 
$\beta$ instead of $\lambda$, with
$\beta  = -\lambda$. We then systematically convert the published $\beta$ into
$\lambda  = -\beta$.

- {\it Units for mass and radius (see also "hints" below):}\\
The default options for planet mass and  radius units are the Jupiter mass and radius. 
In case a referenced paper gives the mass and radius in Earth units, the catalogue automatically converts
them in Jupiter units with the convention 1 $M_{Jup}$ = 317.83 $M_{\oplus}$ and
1 $R_{Jup}$ = 11.18 $R_{\oplus}$ (Allen 1976). The user can change units by clicking on "M$_{\rm Jup}$" 
and "R$_{\rm Jup}$" in the tables (Fig 4.).

- {\it Year of discovery: }\\
The purpose is not to establish a priority among discoverers. 
It indicates  the year of announcement in a professional meeting or the
date of submission of a discovery paper. The date of publication is sometimes the
year after the date of submission or announcement in a professional conference.

The notion of "year of discovery" is problematic for a few objects.   $\gamma$ Cep b   
was more or less strongly suspected as a candidate in 1988 (Campbell et al. 1988). $\beta$ Gem b 
was strongly suspected in 1993 by Hatzes \& Cochran (1993). $\gamma$ Cep b was retracted in 1992 by Walker et al.
(1992)
 and finally reconfirmed in 2003  with the correct mass and period (Hatzes et al, 2003). 
$\beta$ Gem was confirmed in 2006 (Hatzes et al.). For these two objects
we have chosen the date of final confirmation.
HD 114762 b ($\sim$12 $M_{Jup}$) was discovered as a confirmed companion in 1989 (Latham et al.), but it was 
not baptized  as a planet at that time.

 - {\it Status: }\\
R = refereed paper (accepted or published), S = submitted paper, C = announced in a professional conference, 
W = announced on a professional website.

- {\it Errors: }\\
The quoted errors are those given by the discovery paper {\bf or 
subsequent papers based on new observations}. They generally refer
to 1 $\sigma$ errors, with a few exceptions that  are not always clearly documented
in the literature. They are therefore only indicative.
When a paper gives both the statistical and the systematic error, we arbitrarily
 {\bf add the two errors in quadrature}. Sometimes most recent papers give 
a larger error than previous ones, based on 
a deeper analysis. We then take the most recently published error even if larger.
The most important is that the reader can trace back the data through the referenced papers.

- {\it Unconfirmed and retracted candidates:}\\
This table stores candidates 
waiting for confirmation by later measurements. It includes definitely retracted planets
 for users who want to understand the reason of retraction. Details are 
given in the individual pages for each candidate. 

Past experience shows that since the announcement of the not confirmed companion
 Lalande 21185 b (Gatewood 1996), only about 1.5\% of candidates have been retracted. See also
{\rm http://obswww.unige.ch/}$\sim${\rm naef/planet/geneva}\_{\rm planets.html}
\#{\rm SPECIALCASES}
for special or spurious cases.
\subsection{Some hints}

- {\it Sorting}:\\
By default, the tables 1 to 7 sort the planetary systems by increasing period of the planet 
closest to the star. The user can sort the planets by increasing or decreasing values
of each parameter by clicking on the parameter name at the top of each column.

- {\it Extended tables}: \\
The default option of tables 1 to 7 gives planet data only. By clicking on "{\sc more data}"
the stellar parameters appear also (Fig 7a), as well as the calculated value "ANG. DIST." for $a/D$ ($D$ =
distance of the star- Fig 7b.); it gives a first evaluation of the star-planet angular separation. 
In a future
version, we will calculate online the angular separation at maximum angular elongation
for eccentric orbits for a given orbital inclination.

\begin{center}

\includegraphics[width=7cm]{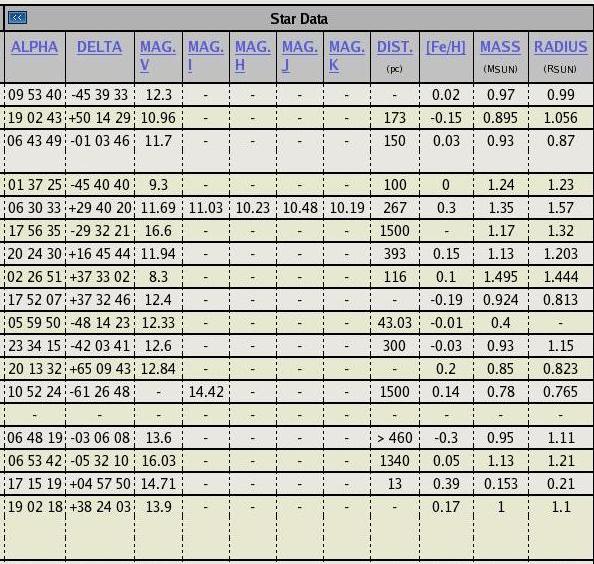}

{\bf Fig. 7a.} Stellar characteristics in the "extended table"

\includegraphics[width=7cm]{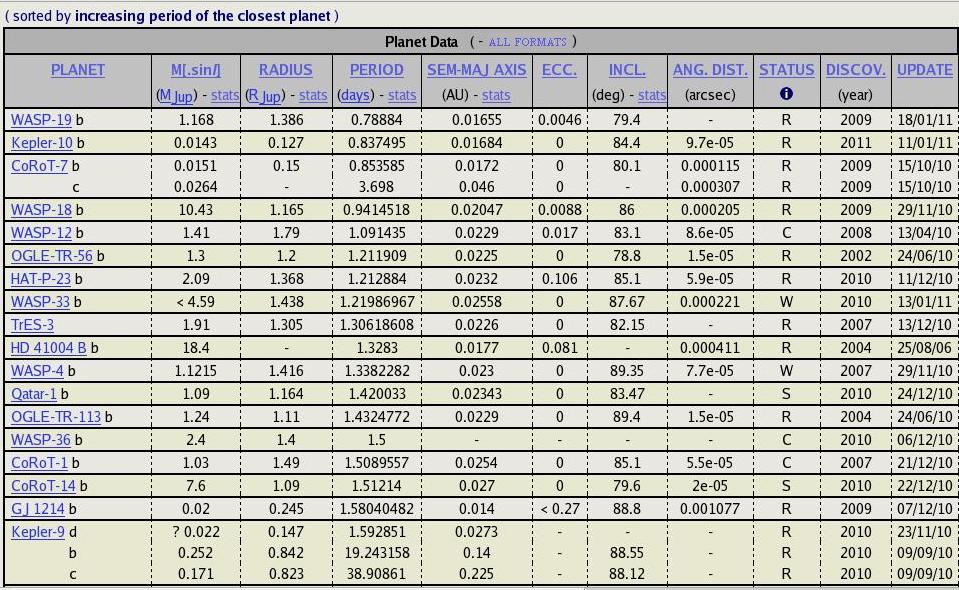}

{\bf Fig. 7b.} Planet table with "ANG. DIST."
\end{center}

- {\it  Planet names}\\
For single planetary companions to a host star, the name is generally 
{\it NNN b} where {\it NNN} is the parent star name. Since all stars have multiple names,
we choose the name as given in the discovery paper: e.g. 51 Peg b instead of HD 217014 b.
For planets with an alternate star name, the user can retrieve the planet through the star page at SIMBAD
which has a link to the present catalog (see Fig. 8. for $\beta$ Gem,  alias HD 62509).

\begin{center}
\includegraphics[width=7cm]{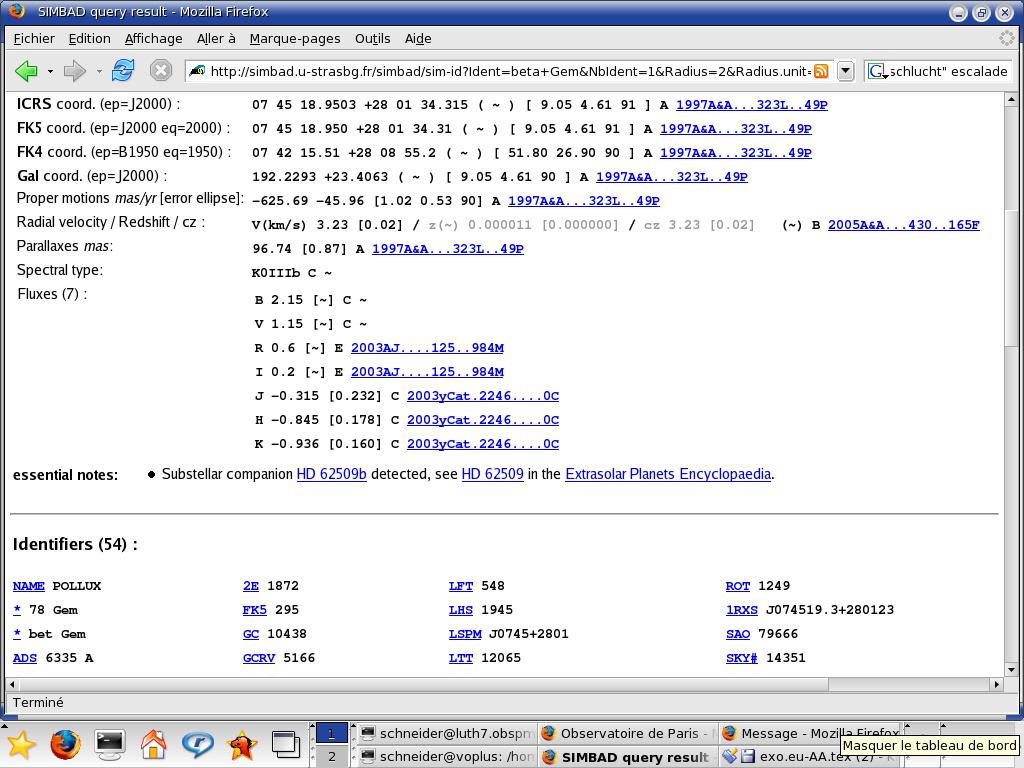}

{\bf Fig. 8.} Simbad page for $\beta$ Gem.
\end{center}

For multi-planet systems, the planet names are {\it NNN x} where 
{\it x = b, c, d,} etc. refers to the chronological order of discovery of the 
planet or to the increasing period for multiple planets discovered at the same time.
Exceptions are possible for planets detected by transit like CoRoT, 
HAT-P, KEPLER, OGLE, Qatar, TrES,  WASP and XO planets or planets detected by microlensing 
like MOA and OGLE, which are based on the name of the 
discovery facility. For some planets  we have arbitrarily shortened 
the name in the tables
for aesthetic reasons: e.g. 2M1207 b instead of 2MASSWJ1207334-393254 b. 
The full name is then in the individual pages.

Another configuration deals with planets circum-orbiting a binary star. 
It is the case of the binary NN Ser. We have followed the discoverer's designation
NN Ser(ab)c and NN Ser(ab)d for the two companions (Beuermann et al. 2010). 
We follow the same type of designation for other candidates circum-orbiting a binary star.

For "free floating" planets, the name is the name given by the discoverers.
\section{Interactive online tools}
\subsection{Statistics and output tables}
The database provides online histogrammes and correlation diagrammes. 
Click on "Histograms" and "Correlation diagrams" at the top of each of the 8 tables.
Some filters on data are provided, as well as the choice between linear scale and log scale.

They do not guarantee a solid scientific value since the biases, resulting in heterogeneous data,
are not well documented in the literature. 
But  they provide general trends and are useful for public presentations.
 We  nevertheless note
an interesting feature revealed by the $R_{pl}$-$a$ correlation diagramme: 
 the scatter
in $R_{pl}$ for giant planets 
is reduced at large distance from the star. This is a first qualitative confirmation that the planet inflation
is due to some influence
of their nearby parent star, heating, tidal effects or whatever (Baraffe et al. 2010). See Fig 9.
It shows that a quick inspection of histogrammes and correlation
diagrammes can have at least some scientific meaning.

\begin{center}
\includegraphics[width=7cm]{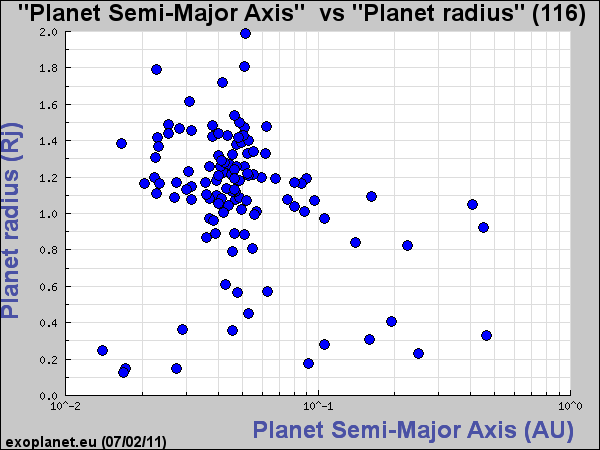}

{\bf Fig. 9.} Planet radius versus semi-major axis.
\end{center}

Output tables are provided in  XML and  CSV format. Click on "Planet Table"
at the top of each of the 8 tables.
\subsection{VO services}
The first Virtual Observatory (VO) service implementing IVOA {\it
ConeSearch}\footnote{{\rm http://www.ivoa.net/Documents/latest/ConeSearch.html}}
interface for positional queries was introduced in the database in 2006. 
Since then it is possible to  locate
and query the web service from the popular VO client applications like
{\sc cds aladin}\footnote{{\rm http://aladin.u-strasbg.fr/}} or {\sc
topcat}\footnote{{\rm http://www.star.bris.ac.uk/~mbt/topcat/}} (see
Fig.~ 10)
%\label{fig:topcat_sphere}) 
by querying VO registry with catalogue
keywords (e.~g. ``exoplanet''). The endpoint URL of this web service
{\rm http://voparis-srv.obspm.fr/srv/scs-exoplanet.php} leading to an .xml file
which can be processed by any standard VO tool.

\begin{center}
%\begin{figure}[]
\includegraphics[width=7cm]{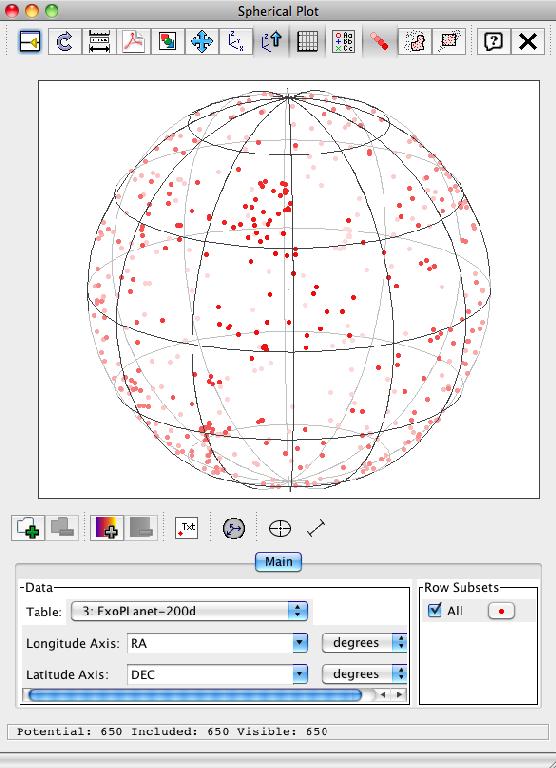}
%\caption{Exoplanet catalogue plotted on the sky in {\sc topcoat}.}
%\label{fig:topcat_sphere}
%\end{figure}

{\bf Fig. 10.} Exoplanet catalogue plotted on the sky in {\sc topcat}.
\end{center}

A web application also provides {\sc tap} (Table Access
Protocol) services, which, being a successor of IVOA ConeSearch
protocol, enables the user to query the dataset with arbitrary filters
either from graphical clients or using endpoint URL
{\rm http://voparis-srv.obspm.fr/srv/tap-exoplanet/} and custom
software client optionally developed by users (see the Appendix).

All the present description of the database is summarized in the README.html file which 
will be updated to account for future evolutions.
\section{The Extrasolar Planets Encyclopaedia}
The database is part of the Extrasolar Planets Encyclopaedia available at 
{\rm http://exoplanet.eu}.
It has been designed since 1995 to encourage and facilitate the development of all exoplanet activities
and communication between researchers.
It gives the latest news, access to online tutorials and general papers,  a list of 
current and projected ground and space searches for planets, an extended bibliography, 
a list of past and future meetings, links to theory work and to other sites (Fig 11).

The bibliography gives more than 8000 references (from Epicurus to today): articles in professional journals
and preprints, books, conference proceedings, PhD theses. It is updated daily and
can be queried directly by author names, paper titles or both:\\
{\rm http://exoplanet.eu/biblio-search.php?Authors=name}\\
{\rm http://exoplanet.eu/biblio-search.php?mainWords=title}\\
{\rm http://exoplanet.eu/biblio-search.php?mainWords=title}
{\rm \&Authors=name}

\begin{center}
\includegraphics[width=7cm]{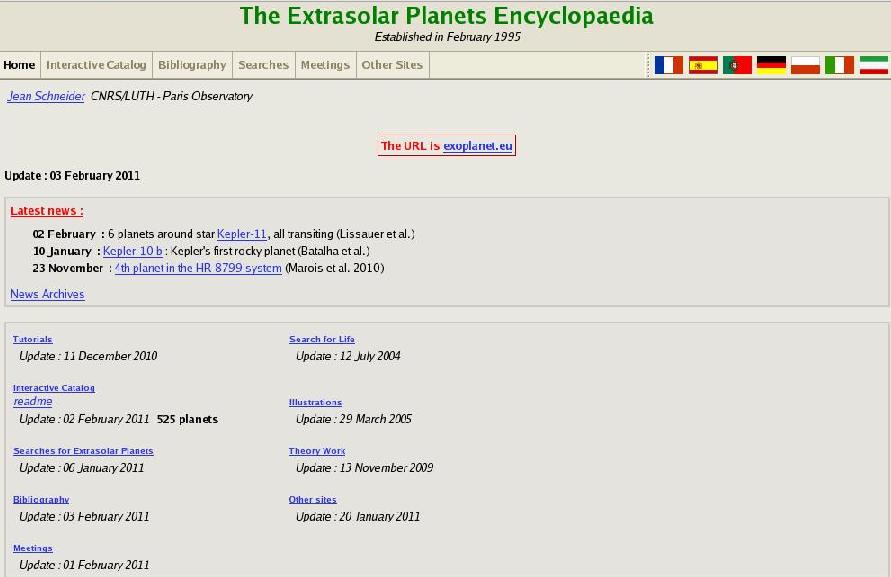}

{\bf Fig. 11. Home page for the Extrasolar Planets Encyclopaedia}
\end{center}

\section{Future developments}
The database will be upgraded continuously in several aspects: 
addition of new planets, addition of new data for each planet,
addition of new links and services.

- {\it New data:}\\
 We will add several new planet characteristics such as the position angle, number of planets in multiple systems, spectra, albedos, planet calculated and measured temperature, rings, moons, etc.

By anticipation of the discovery of  exomoon-like companions (and possibly
binary planets (Cabrera \& Schneider 2007)), which can happen any time now by transits,
we propose the following solution for their naming:  NNN b1, b2 etc, if they have a similar 
semi-major axis and if the separation between the
companions b1, b2 etc is permanently less than the Hill radius (in order to
make a distinction with other types of 1:1 resonances  like  exchange orbits (Funk et al. 2010), 
eccentric resonances (Nauenberg 2002) and Trojan
planets (Dvorak et al. 2004)).

- {\it New links:}\\
 Links to NStED and Exoplanet Data Explorer individual pages for planets, links to data tables at CDS.

- {\it New services and VO aspects:}\\
 We are preparing the management of multiple star names, multiple filters, etc.
For the VO aspects of the database, we are preparing  a new version
of web applications distributing the catalogue.
We plan to implement an advanced cross-platform client toolkit for easy
intercommunication with arbitrary VO applications by means of {\sc
samp}, Simple Application Messaging Protocol. The goal is to make a
web browser act like a data browser which helps users to locate datasets
they need and send it flawlessly to dedicated VO tools 
launched before in a background, where in turn all scientific analysis
takes place. This significantly enriches the user interaction with the
data adding an opportunity to do sophisticated scientific analysis
online. 

To conclude, having two or more independant catalogues allows to have 
complementary services and each reader can check their mutual consistency.
Comments and questions on the database can be addressed to Jean.Schneider@obspm.fr.
\begin{acknowledgements}
We thank F. Arenou, S. Ferraz-Mello, T. Guillot, T. Mazeh and G. Wuchterl for discussions, 
J. Normand and C. Cavarroc for technical help, {\bf J. Cabrera, H. Deeg, K. Gozdziewski,
M. Heydari-Malayeri, I. Pagano
and N. Santos for their translation in their languages,}
and the users, {\bf in particular Danilo M. Bonanni and Anthony Boccaletti,} who have pointed out typos 
or provided useful comments. {\bf We also thank the referee for useful suggestions}.\\
\end{acknowledgements}

\bigskip

\begin{appendix}

\begin{center}
APPENDIX\\
\end{center}

 There are two ways of access The Exoplanet Encyclopaedia data using VO
 protocols. First one is manual access through a web browser using
 VO-Paris Data Centre Portal at
 {\rm http://voparis-srv.obspm.fr/portal/} and interfaces available for
 data discovery therein. Another option is to employ one of the
 available client applications. We will give a brief step-by-step
 explanation by example of TOPCAT tool.

 To access the exoplanet data one has to undertake the following steps:

 \begin{enumerate}

 \item Launch TOPCAT by opening the link to its Java WebStart version
 at {\rm http://andromeda.star.bris.ac.uk/~mbt/topcat/topcat-lite.jnlp}.
 There must be a Java installed on the user's computer, including WebStart
 support in a browser. All modern operating systems now support this
 mainstream technologies.

 \item Go to {\it File} $\longrightarrow$ {\it Open}, then in a raised
 window click on {\it DataSources} $\longrightarrow$ {\it Cone Search} and
 type ``exoplanet catalog'' in the {\it Keywords} field of the last
 window opened and press {\it Submit Query} button. In this step the
 user has access to all the services related to exoplanet catalogs available in
 the Virtual Observatory, searching yellow pages analog, called
 registry.

 \item Click on the line with "ExoPlanet" label in the {\it shortName}
 (first) column in the query results section to select it. This
 indicates that user is going to access The Exoplanet Data
 Encyclopaedia ConeSearch service.

 \item Use the form below to fill the details of your coordinate
 request. To load the whole catalog, for example, use query with RA=0,
 Dec=0 and Radius=360. Press {\it OK} to retrieve the data.

 \item To visualize the result, one can plot the all-sky distribution
 of the objects in the catalog by choosing {\it Graphics}
 $\longrightarrow$ {\it Sky} item from the main menu.

 \item For any other kind of plots go to {\it Graphics} menu and choose
 appropriate graphical representation from the list of available ones.
 With these options it is possible to plot any column of the data
 table.

 \end{enumerate}

The Virtual Observatory gives the ability to cross-correlate any
quantities between different catalogs by using UCDs, Unified Content
Descriptors, which associate same physical parameters and their units
with each other.

 As an example of interoperability with other VO resources, one can
 plot another exoplanet catalog over the loaded one. To do so:

 \begin{enumerate}

 \item Make step 2 from the previous instruction.

 \item Choose the line with "EXOPLANETS" label in the {\it shortName}
 (first) column. It is taken from the NASA/HEASARC query tool
of the Extrasolar Planet Encyclopaedia at \\
{\rm http://heasarc.gsfc.nasa.gov/cgi-bin/W3Browse/w3query.pl?tablehead=name=\\
heasarc\_exoplanets\&Action=More+Options\&Action=Parameter+Search\&ConeAdd=1}

 \item Make step 4 from the previous instruction, then select first
 table from the {\it Table List} on the left of main TOPCAT window and
 then make step 5 from the previous instruction.

 \item Press {\it Add a new dataset button} in the toolbox below the
 graph with all-sky distribution.

 \item Select table 2 in the drop-down menu {\it Table} in the {\it
 Data} section of the newly opened tab.

 \item Click on the blue marker point on the right of the graph window
 in the section {\it Row Subset}.

 \item In the raised window of marker properties, choose open circle
 for a marker in a drop-down menu {\it Shape} on the left and increase
 its size till 5 in the drop-down menu {\it Size} on the right. Press
 {\it OK} to apply changes.

 \item Now you have 2 exoplanet catalogs overplotted on a sphere. You
 can move the sphere by holding your left mouse button and moving the
 mouse and zoom/unzoom it with your mouse wheel.

 \end{enumerate}

\end{appendix}

\end{document}